\documentclass[conference]{IEEEtran}
\IEEEoverridecommandlockouts

\usepackage{cite}
\usepackage{amsmath,amssymb,amsfonts}
\usepackage{algorithmic}
\usepackage{graphicx}
\usepackage{textcomp}
\usepackage{xcolor}
\usepackage{subfig}
\usepackage{url}

\def\BibTeX{{\rm B\kern-.05em{\sc i\kern-.025em b}\kern-.08em
    T\kern-.1667em\lower.7ex\hbox{E}\kern-.125emX}}
\begin{document}

\title{Developing Medical AI : a cloud-native audio-visual data collection study}

\author{\IEEEauthorblockN{Sagi Schein, Greg Arutiunian,
                                            Vitaly Burshtein,Gal Sadeh,\\
				 Michelle Townshend, Bruce Friedman, Shada  Sadr-azodi}
\IEEEauthorblockA{\textit{GE Healthcare}\\
Email: 
sagi.schein@ge.com,
greg.arutiunian@ge.com,
vitaly.burshtein@ge.com,\\
gal.sadeh@ge.com,
michelle.townshend@ge.com,
bruce.friedman@ge.com,
shada.sader-azodi@ge.com
}}

\maketitle

\begin{abstract}
Designing Artificial Intelligence (AI) solutions that can operate in real-world situations is a highly complex task. Deploying such solutions in the medical domain is even more challenging. The promise of using AI to improve patient care and reduce cost has encouraged many companies to undertake such endeavours. For our team, the goal has been to improve early identification of deteriorating patients in the hospital.  Identifying patient deterioration in lower acuity wards relies, to a large degree on the attention and intuition of clinicians, rather than on the presence of physiological monitoring devices. In these care areas, an automated tool which could continuously observe patients and notify the clinical staff  of suspected deterioration, would be extremely valuable. In order to develop such an AI-enabled tool, a large collection of patient images and audio correlated with corresponding vital signs, past medical history and clinical outcome would be indispensable. To the best of our knowledge, no such public or for-pay data set currently exists. This lack of audio-visual data led to the decision to conduct exactly such study. The main contributions of this paper are, the description of a protocol for audio-visual data collection study, a cloud-architecture for efficiently processing and consuming such data, and the design of a specific data collection device. 
\end{abstract}


\section{Introduction}
In recent years Artificial Intelligence (AI) has grown from being an academic curiosity into a powerful methodology that can solve complex real-world problems, such as autonomous driving, predicting consumer purchasing decisions and even automatically translating foreign languages. In specialized tasks such as, automated face recognition and some board games (Chess, Go), AI has already surpassed human level accuracy. The most prominent AI methodology, these days, employs large neural networks that are trained with correspondingly large data sets, colloquially termed Deep Learning (DL). Training these large neural networks requires massive amounts of data and powerful compute infrastructure; collecting these large data sets might turn into a complex and an expensive operational process. In the medical domain, there are additional factors which make this kind of projects even more complex. These include working in the hospital environment to collect patients data, adhering to strict regulatory and privacy practices and the need to formally prove safety and efficacy of the resulting model. The promise of utilizing AI to improve patient care, however, has attracted many companies to undertake such endeavours. The goal of our team has been to create AI tools that could assist clinicians in identifying patients at risk of deterioration, by using video and audio signals.

 Patient deterioration is a dynamic process which may exhibit high variability. In other words, every patient might have a unique deterioration path that is difficult to predict. Patients may deteriorate in any care area in the hospital. In the lower acuity wards, the increasingly high patients-to-clinicians ratio may result in deterioration cues going unnoticed until it is too late. In many such cases, the watchful eye of a trained clinician can potentially identify at-risk patients, raise a notification and hopefully save lives~\cite{Worry, Color}. To identify deteriorating patients, the clinician must synthesize changes to the patient appearance, mental state, breathing patterns and pain levels over time, in combination with their medical context and vital signs. Being able to capture these capabilities in an AI tool could clearly be useful. Development of such a tool will require a large collection of patient audio-visual data, correlated with their vital signs and clinical outcome. Due to privacy, security and cost limitations, no public or for-pay data sets currently exist. 

The main contribution of this paper is a blueprint for conducting an audio-visual data collection study in the cloud. This blueprint includes the outline for a custom-built data collection device, the architecture of a cloud-native efficient data processing pipeline, and the study protocol.

The paper is organized as follows,  in Section~\ref{sec:data_col_infra}  the data collection software and hardware infrastructure is described. Section~\ref{sec:rdb} details a data consumption layer which is optimized for training AI model.  Section~\ref{sec:clinical_study} describes how the first phase of the study was operationalized in a simulated hospital environment and discusses the requirements for a large-scale audio-visual data collection study in a hospital. In Section~\ref{sec:results} initial results from the first phase of the study are given. The paper concludes and suggests some future directions in Section~\ref{sec:conclusion}.

\section{Data Collection Infrastructure}
\label{sec:data_col_infra}
\begin{figure*}[t]
\includegraphics[width=\textwidth]{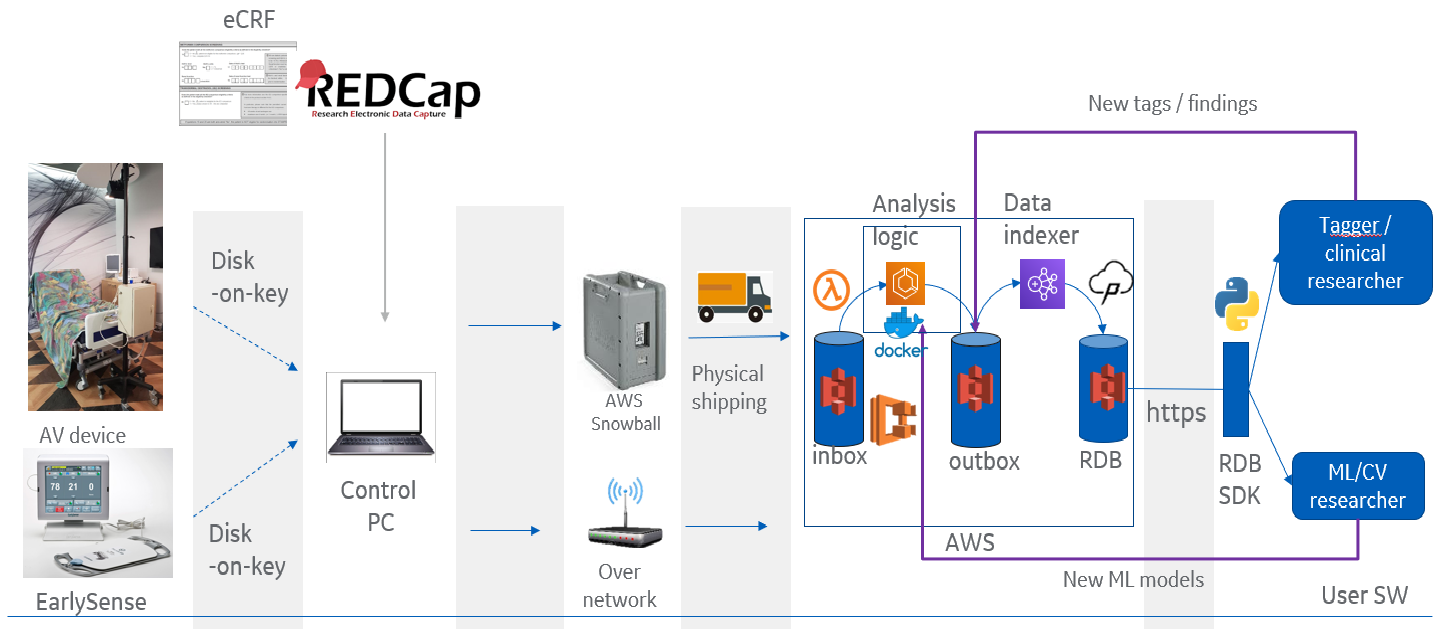}
\caption{The study data processing pipeline begins on the study site where data is collected with the collection device into portable encrypted storage The data is then transferred to the cloud, ingested and indexed for supporting AI research use cases. The pipeline is a cloud native application which utilizes several Amazon Web Services (AWS) services to create a scalable and economical solution.  }
\label{fig:curation_pipeline}
\end{figure*}
Several requirements were defined at the onset and when constructing the infrastructure to support this study. First, collecting data in a hospital ward requires the study to have minimal impact to the patient care, as well as, minimizing additional burden on hospital staff. Second, the study should have minimal reliance on hospital IT and/or the biomedical department. Third, local legal requirements may impose that raw identifiable and personal data might have to be stored in the country of origin, whereas the researchers who would use it could be located elsewhere. Lastly, the size of  the collected data can be very large, for example, during a ten-week study period, more than two terabytes of data were collected in an initial phase of the study (described in Section~\ref{sec:clinical_study}). This data must be curated, analyzed and stored efficiently and economically. To support the aforementioned requirements, it was decided to design and build a dedicated collection device which could perform effectively in a clinical study setup and deploy all data processing and storage in the cloud. Participants' data would be collected in the hospital and sent over secured communication channels to the cloud infrastructure. Once curated, enriched and organized the data becomes available for designing AI algorithms while adhering to data management standards (GDPR and HIPAA compliance standards) for encryption and access monitoring.

Figure~\ref{fig:curation_pipeline},  illustrates the general flow of data in the study. The left side of the figure shows the audio-visual collection device, labeled henceforth AV-device, which is described in Section~\ref{sec:data_collection_hw} and the vitals collection device, a piezoelectric device from EarlySense~\cite{EarleSense}. Both devices store their data to an internal storage and are periodically copied, through a control PC, to the cloud. This can be done either through the hospital network or using a portable data storage device~\cite{Snowball}. Once in the cloud, the data is automatically curated and enriched with a sequence of automatic algorithms, as described in section~\ref{sec:processing_pipeline}. It is, then, organized into a research database (RDB) for consumption by AI researchers. RDB may serve as a data source for multiple applications such as, tagging applications, data analysis tools and for training machine learning models. The design of RDB is outlined in section~\ref{sec:rdb}. 

\subsection{Data collection device}
\label{sec:data_collection_hw}
In order to collect audio-visual data in the hospital environment a dedicated collection device was constructed. The operational parameters of the device included the ability to collect visual data and audio of patients in bed, at a constant rate, from a distance of up to three meters. The device was required to work independently for 24 hours without constant human monitoring. The device had to operate independently of the hospital utility infrastructure i.e. power and communication network. Furthermore, the design had to protect the privacy of the participants such that their images and audio could not be compromised. 

\begin{figure}[htbp]
\centering
\includegraphics[width=1.4in]{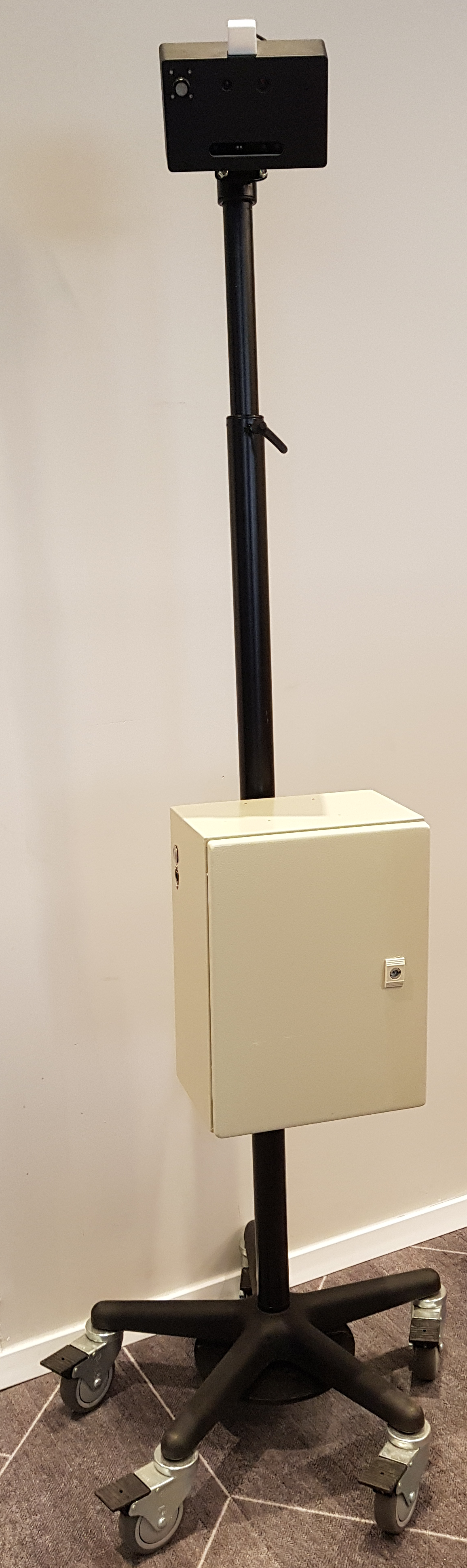}
\includegraphics[width=2in]{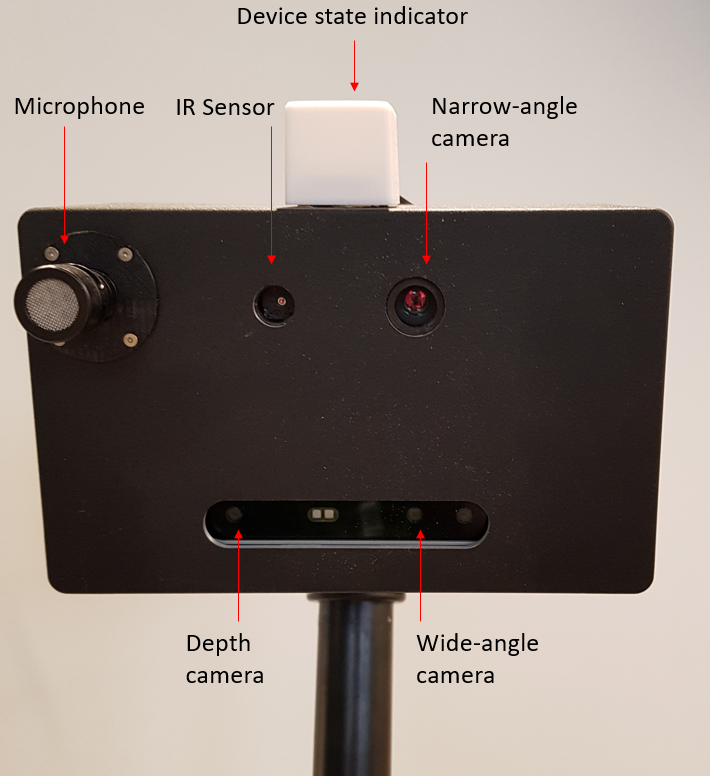}

\caption{$(a)$ the data collection device. The upper part shows the sensor head while the middle cabinet holds the device compute, storage and batteries. $(b)$ shows a close up of the sensor head.}
\label{fig:device}
\end{figure}
   
Inspecting patient's behavior, motion, and interaction with others requires a wide-angle camera capable of monitoring the bed and surrounding area. At the same time, existing evidence suggests that important information regarding the patient's  state can be gleaned from  their facial expression, neck and abdominal muscle contraction, and breathing patterns. As such, high fidelity images of the upper part of the body are necessary. To satisfy both  requirements  the device was fitted with two simple cameras with fixed fields of view, wide  and narrow angle. To improve the accuracy of scene analysis techniques, a depth camera was also integrated. Due to its small form-factor, ease of integration and good depth accuracy Intel RealSense D415~\cite{RealSense} was selected for both depth and wide-field channels. This camera allowed up to 25 frames per second at a 1280x720 resolution for both the color and depth channels. For a narrow angle color camera, Sony IMX219 sensor with a 4 to 1 optical lance, was  selected.  

Remote sensing of patient's distribution of skin temperature may potentially indicate multiple medical conditions, such as fever and possibly changes to the patient blood circulation. An infrared (IR) camera, was integrated in the device, to enable the detection of patient motion and the construction of heat distribution of the patient and surroundings, even during the night under dim lighting conditions. Based on performance to cost trade-off analysis the Lepton 3.5 sensor module from FLIR~\cite{FLIR}  which has  spatial resolution of 160x120 heat pixels, and a frame rate of 8 frames per second, was selected. The sensor does not require cooling or calibration on site, both of which were important in the study setting. 

To collect patient audio, the device was also fitted with a narrow angle microphone that is capable of capturing noises and speech at approximately three meters. Care was taken to ensure that the microphone had good directionality, meaning it could collect sounds from a narrow cone directed at the bed. Such a narrow angle helped to reduce background interference, allowed algorithms to focus on the correct sounds and, may also help protecting the privacy of caregivers and other nearby patients.

Figure~\ref{fig:device}$(a)$ exhibits the data collection device in the lab, Figure~\ref{fig:device} $(b)$ demonstrates a close up image of the sensor head. On the top of sensor head there is a light indicator which reports the device operational state. The microphone is located on the left side of the sensor head, the IR camera and the narrow angle camera are in the middle while the RealSense depth and wide-angle cameras are shown at the bottom. 

\begin{figure}[htbp]
\includegraphics[width=1.5in]{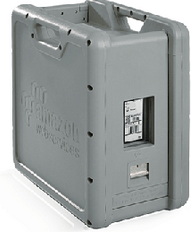}
%\put(-10,10){$(a)$} to do
\caption{The AWS Snowball data storage utility.}
\label{fig:snowball}
\end{figure}

To simplify the operation of the data collection device,  it was decided to only include two buttons: a power button and a privacy button. The power button turned on the device and enabled a data collection mode. The privacy button allowed clinicians and patients to pause data collection when ever necessary.  At the end of the day, the power button was used to power down the device and verify that all the data is stored securely. To configure the device, an external laptop or tablet was connected to its graphical user interface (GUI) allowing a study coordinator to enter, for example, the participants ID, the ward ID and the device ID. During operation, the device  stored all collected data to an encrypted external Solid State Disk (SSD) which was locked in the device cabinet. The same cabinet also housed the device battery. The battery was capable of supporting 24 hours of device operation. Each device was equipped  with a pair of batteries and a pair of hard disks. During the day, used-up batteries were recharged while the stored data was transferred to the cloud or to a Snowball device (see Figure~\ref{fig:snowball})~\cite{Snowball}.

A data collection device generates approximately 7 gigabyte of images and audio per hour. To upload the data to the cloud, two alternatives were designed. For small studies (e.g. less then four devices per site), a network upload procedure was provided. This involved the study coordinator to remove all external disks from the devices and to connect them to the control PC. A dedicated upload software was then used to upload the data to the cloud. The benefit of this approach was that it allowed the data to be analyzed continuously, inform the study coordinator of any quality issues and rapidly react to failures. Another benefit of this approach was that it reduced the peek load on the data curation pipeline. 

For large scale studies, where a network upload is infeasible, AWS Snowball~\cite{Snowball} could be used as an alternative. AWS Snowball is a physical data transfer utility from Amazon which may store up to 100TB of data on a single chassis. It is designed to be shipped to the customer's site where it would be loaded with historical data archives then returned to AWS data center. In a large scale study, AWS Snowball could be re-purposed as a storage utility which would be filled daily with audio, video and physiological data. The approach would allow a study to be totally decoupled from the hospital IT infrastructure and would reduce the limitation on the size of the data captured during the study resulting in additional operational complexities.  The downside of this approach is that it may create a single point of failure in the data management pipeline. For example, if a Snowball device gets lost on-route, data would be lost. Having a second of AWS Snowball on site and storing two copies of the data could significantly reduce this risk.

\subsection{Data curation pipeline}
\label{sec:processing_pipeline}

To efficiently curate and enrich the data, a cloud-native curation pipeline was designed and constructed. Cloud computing platforms offer many services that can be combined to form robust and scalable systems and reduce their operational costs even. Due to commercial consideration the data curation pipeline was deployed in the AWS cloud, nevertheless, equivalent services exist in most cloud computing platforms.
There were several functional requirements that needed to be addressed to successfully construct the data curation pipeline. First, since collection of clinical data is expensive, it becomes imperative to minimize the probability of data loss while it is being processed. Therefore, AWS S3~\cite{S3}, a distributed, redundant and resilient key-value storage was used. In addition, it was decided that raw study data is always written and never modified which reduces the probability of inadvertent data losses. To further reduce this risk, a second copy of the data was placed into AWS Glacier~\cite{Glacier} a low-cost archiving service. A second functional requirement recognized that there could be huge deviations in the data upload rate. Data could be slowly uploaded over the network for a few hours each day or it may arrive directly from the AWS network backbone when a Snowball device is connected. For the remainder of the time, the system would be in idle mode, waiting for data to arrive. The proposed system architecture had to be able to scale up so that every image or video file could be processed regardless of the load and, at the same time, incur very little idle time costs. 
To this end, the system used AWS Lambda as a scalable front-end service. When a file was copied into the study S3 bucket, its name was added to an entry queue. Due to its robustness, simplicity and cost characteristics,  AWS Simple Queue Service (SQS)~\cite{SQS} was selected for this task. SQS served as a buffer which can cope with any expected data load in the study. Once in the queue, a format conversion lambda function transformed each file to a common file format, for example,  TIFF to PNG  and WAV to FLAC for image and audio files, respectively. A second lambda function was then called to generate a metadata item for each file which was placed in a processing queue for further analysis. This metadata item contained the location of the data item and an empty template of all the features that would be computed for that data item at the later stages of the curation pipeline.

AWS Lambda can be very effective for processing large amounts of data in parallel. It is fully managed and can scale horizontally to thousands of processing functions with zero burden for the end-user. The caveat is that AWS Lambda may become expensive on lengthy compute operations and, more importantly, it cannot use a Graphical Processing Units (GPU) to accelerate compute operations. 
For the designed feature extraction procedures, using a GPU proved to be 10 to 60 times faster than computing them on the CPU, making AWS lambda exceedingly expensive. Instead, all feature extraction computations were implemented as a Docker~\cite{Docker} container and deployed to AWS Elastic Container Service (ECS)~\cite{ECS} where GPUs are available. 

There may be many features useful in predicting patient deterioration. For example, the analysis of changes in patient position and  movement patterns and activities over time could indicate changes in the patient's level of consciousness. Additionally, changes in the patient's skin color may indicate changes in tissue oxygenation and perceived health~\cite{Color} and changes in facial expression may directly hint to patient deterioration~\cite{Faces}. Other important features, such as changes in the frequency and type of cough may be an indication of a respiratory distress. In the data curation pipeline, such features were computed for every data item. These features were later used to create machine learning models. As this research evolves more features could be extracted and enrich the data set. Once matured, such algorithms could become part of real-time clinical AI solution for improved patient care. The description of features that are computed, the algorithms that are used for model training and the clinical goals for which they are designed are out of the scope for this paper and will be published in future research papers.

% 
%When deploying AI tools in real environments, it is usually beneficial that underlying models could continuously evolve and adapt to actual data. The medical setup, however, imposes strict regulatory constraints on using patients' private data, such as face images and sounds. In a controlled clinical study these restrictions would be handled by ensuring participants' consent prior to collecting their data. In a real hospital environment getting such a consent could prove  much more challenging. To mitigate this difficulty a two phase approach was proposed. In the first phase, a set of DL algorithms would extract low-level features from controlled data sets, i.e data sets that were acquired in studies or, when available, from consented patients. These features effectively obfuscate subjects' identity and preserve their privacy. The second phase would utilize these features to train a high-level AI models which have well-defined clinical goals. The algorithms in this phase could be trained on features that is acquired in the hospital as part of the system's normal operation and transformed by the feature extraction algorithms as well as directly using study data. 
%Using feature data to train high-level models would perhaps enable the use of flexible consent forms similar to those hospitals already use these days when asking patients to share generic medical data.  The caveat of such two-phase approach might be a reduction in flexibility and model accuracy.

\section{Research database}
\label{sec:rdb}
The result of the curation and enrichment pipeline are per-file metadata objects. These include, the location of persons' head/neck and their limbs, their facial expression encoded as facial action units~\cite{FACS}, and their activities of daily life ( sitting, standing, laying down, eating).  In addition, the data contains heart rate (HR) and respiratory rate (RR) measurements, a classification of the subject's medical condition and some manual tagging of parts of the data by subject matter experts (SME). For instance, imagine a researcher who needs to develop an AI model for predicting the severity of Chronic Obstructive Pulmonary Disease (COPD) from facial expressions. To build such a model, the researcher would have to select a cohort from the data of all patient images who suffer from COPD,  are looking towards the camera and have valid action unit encoding. Sifting through tens or hundreds of millions of images just to create such a cohort could be time consuming and expensive. Every time the researcher changes the definition of the cohort, the same procedure would need to be repeated. A much better approach would be to structure these search criteria into a database query and let the database system do the heavy-lifting for us. The caveat is that relational databases were never built to handle datasets of 10's of terabytes of images and audio. Trying to use them for this capacity requires powerful servers and very fast storage that are always-on and result in an uneconomical solution.  At this  point, one might suggest to store just pointers to the actual data. This could work, as long as the system is located on a fast file system and close to the point where the data is consumed. Unfortunately, when data is in the cloud, the costs of such fast file system is high and it is unsuitable for a long term, robust, storage of large amounts of data. Trying to use S3 for this function would result in many reads of small data items, leading to low throughput of the AI training algorithm.

The solution to these difficulties is to forgo some of the flexibility of a relational database, for example, to optimize the system for reading of large partitions of data, allow new writes to accumulate before they become available for consumption and only have de-normalized data with no join operation. The Research Database (RDB) is a cloud native data-lake system created specifically with petabyte-scale research use-case in mind, and with the flexibility to allow exploration, Business Intelligence (BI) and AI use-cases. It utilizes S3 as its storage layer and Apache Parquet~\cite{Parquet},  a columnar file format for storing data. A data set in RDB can be conceptualized as a table in a relational database. In RDB a row in a table may include meta-data columns together with binary columns which represent images, audio and other binary data types. Having Parquet as a storage format, makes RDB compatible with large variety of query engines like AWS Athena~\cite{Athena} and Redshift Spectrum~\cite{Redshift} which allow query and exploration of meta-data at scale. DL and other AI use-cases are covered by a dedicated SDK that is realized in the RDB-client. 

The RDB-client is implemented as an installable Python~\cite{Python} package which exposes an iterator like interface to RDB data-sets together with select, shuffle, filter, transform, cache and deploy capabilities. Instead of downloading all the data required for learning, the client would stream it to the AI training infrastructure utilizing all available network and CPU resources to fully utilize the GPU. The streamed data may include all metadata and on-the-fly decoded Numpy~\cite{Numpy} arrays.  By supporting online transform functions the RDB client removes the need for pre-processing and storing of multiple copies of data. The RDB client exposes an iterator interface which conforms with most modern AI/DL python frameworks allowing a straight-forward integration. In addition to streaming independent rows the RDB client allows streaming of ordered sequences or NGrams.  In cases where streaming pre-processing is undesired the RDB client utilizes a managed local cache of the streamed data. When enabled, the streamed and transformed data are only created once and then reused in all subsequent computations. The RDB client heavily utilizes the Petastorm~\cite{Petastorm} package which was developed by Uber Advanced Technologies Group (ATG), extending it where needed.

The RDB-client also includes deployment utilities which utilize Docker containers and AWS Sagemaker~\cite{Sagemaker} service to allow responsive local debugging and remote execution of the same code. These utilities simplify the deployment process and lower the cost of training by utilizing managed Spot Training~\cite{Spot} in AWS Sagemaker. The use of Docker makes the deployment flexible and allows it to run AI training sessions in a variety of alternative AWS services as well as on local compute infrastructure. RDB is also highly cost-effective as  it uses no databases or expensive cloud services. Its storage costs are mainly dictated by the cost of S3, while its  training infrastructure utilizes low cost AWS Spot Instances or existing on premise compute resources.

\begin{figure*}[t]
\includegraphics[width=\textwidth]{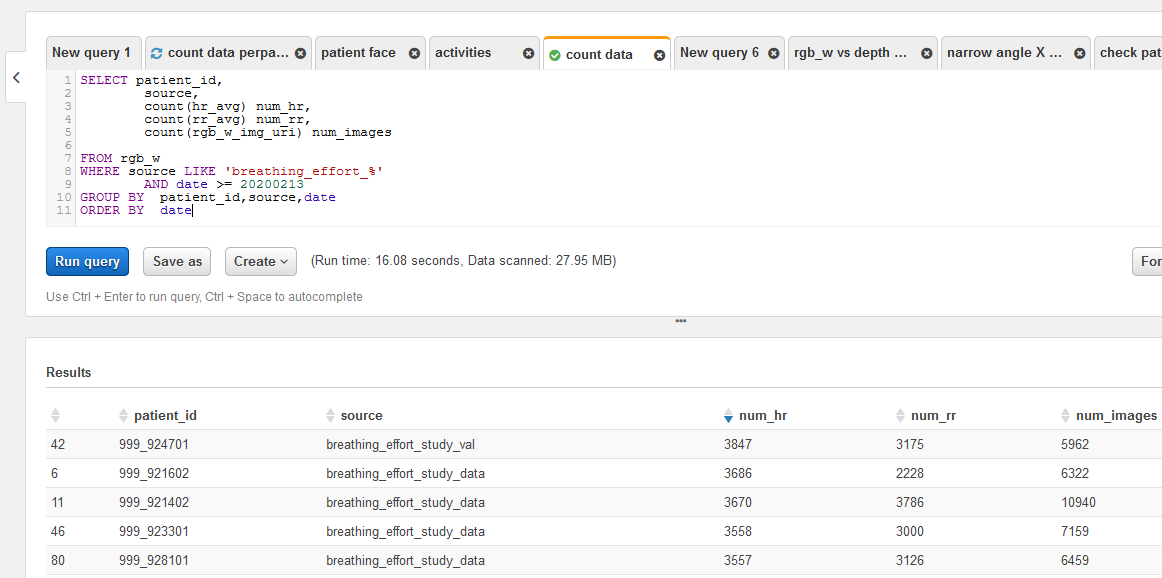}
%\put(-10,10){$(a)$} to do
\caption{AWS Athena may be used to rapidly investigate the study data. The sample SQL query computes the number of physiological samples each subject has and the number of wide angle images.}
\label{fig:athena}
\end{figure*}

\section{Clinical Study}
\label{sec:clinical_study}

Conducting a clinical study in a hospital, may prove time consuming and complicated.  Legal and contractual issues must be resolved before the start of such study. 
In many cases, conducting an initial phase of the study with volunteers out of the formal hospital setting could prove very useful. AI researchers could get valuable information from relatively healthy volunteers which may be used in the development of the algorithms. Additionally, a preliminary phase may also help refine workflows, detecting operational malfunctions before getting into the hospital.  

\begin{figure}[htbp]
\includegraphics[width=3.3in]{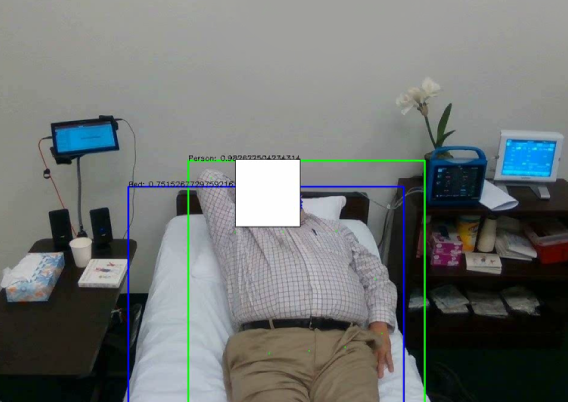}
%\put(-10,10){$(a)$} to do
\caption{A view of the study setup in the simulated hospital room. The participants in this study were people living with one or more chronic conditions.  The face of the participant has been obfuscated to preserve the participant's privacy.}
\label{fig:bi_study}
\end{figure}

In an attempt to obtain surrogate data such initial study was conducted. Its focus was on people with chronic conditions such as COPD, arthritis, and asthma, among other chronic conditions. The participants were recruited from the general population in Chicago and surrounding areas. Participants were pre-screened over the telephone to verify their eligibility and health. Each participant was sent an email with detailed background material regarding the study. Upon arrival to the testing facility,  participants received a thorough explanation about the study and signed a consent form. They were then escorted to the study room were their heart rate, blood pressure (BP) and oxygen saturation (SPO2) readings were obtained. At this point, a study coordinator had an opportunity to disqualify a participant if their BP or SPO2 were outside of the allowable limits. Once all preparations were done, a tablet in the participant room played a pre-recorded set of instructions to the participant. The instructions, as outlined in the protocol were a set of  tasks intended to represent activities of daily living (ADLs), such as, talking on the phone, reading a children's book, simulated drinking, laying down and, sitting on the edge of the bed. Figure~\ref{fig:bi_study}  show the study setup. On the left, there is the tablet which automatically guides the participant throughout the protocol and a bi-directional audio communicator which allows  study coordinator and  participant to communicate. On the right of the bed, an EarlySense~\cite{EarleSense} device, which continuously measures heart rate and respiration rate and a Dinamap~\cite{Dinamap} device, which is used to record SPO2 and blood pressure at the beginning and the end of each session. The blue and green overlaid rectangles show the output of two of the analysis algorithms that are used to enrich the data,  the location of the bed and the location of the subject. At the end of each day, the study coordinator uploaded the audio, images  and physiological  data from each device to the cloud for processing. Once the data was processed, usually with in a few hours, a data verification procedure would sample the data and identify mistakes. These mistakes were communicated back to the study coordinator so that they could be corrected and / or avoided on the subsequent day.

The experience that was gained in the initial phase of the study allow us to structure several guidelines for conducting a data collection study in the hospital. These guidelines are explored herein. Collection of audio-visual  data in a hospital  ward  should interfere with normal patient care as little as possible. To this end, a dedicated study coordinator would work with the hospital staff to identify potential participants in the patient  population.  The role of a study coordinator would be to describe the study to prospective subjects and obtain their signed consent. Upon  patient consent,  the study coordinator would set up the data collection infrastructure.  This would includes  the  data collection  device, described in  Section~\ref{sec:data_collection_hw}, and a physiological sensing device for continuously measuring heart rate and respiration rate e.g. EarlySense~\cite{EarleSense}. The study coordinator would track patients until the end of their study period which may result from the patient discharge,  transfer to an Intensive Care Unit (ICU) or death. To ensure patient privacy,  each admitted patient should be assigned a unique study identification number. This number would be automatically attached to their corresponding image, audio and physiological data files. All other patient information should be locally managed by the study coordinator in two databases.  The first database would manage all the health data that is relevant  to the study, i.e. age, gender, weight,  height,  medical  history  etc. Private data should be obfuscated,  e.g. the exact patient age, weight and height would be discretized to ranges before it becomes available for research purposes.  A second database, which would contain a mapping of study identification number to participants personal identifiable information should be separately managed. This database remains opaque to the researchers and could be used by a study coordinator to manage participants data once the study is completed.

\section{Results}
\label{sec:results}
The data that was acquired in this study was ingested and enriched with automatically generated features. Figure~\ref{fig:athena} shows the result of a sample query to the data. For each participant ID, the query retrieves the number of vital signs samples, HR and RR, and the number of images from the wide-angle camera. A Structured Query Language (SQL) interface makes data exploration highly efficient, allowing researchers to efficiently explore the data and design required cohorts. The same queries can then be transferred to the RDB client and used for streaming them to the AI training procedure.   Table~\ref{tab:study_summary}  depicts some of the summary statistics of the current data-set in the RDB. There were 453 subjects recruited for the study over a 75 day period. Out of the total recruited subjects, 369 subjects completed the study while 84 were disqualified, mainly due to high blood pressure readings. During this time more then 11 million images and audio files were collected which consume roughly 2 Terabyte of storage space in S3.

\begin{table}[h]
  \begin{center}
   \caption{Statistics for the chronic condition study }  	
    \label{tab:study_summary}   
    \begin{tabular}{|l|c|} % <-- Alignments: 1st column left, 2nd middle and 3rd right, with vertical lines in between
      \hline
      Total recruited subjects & 453 \\
      Total completed subjects & 363 \\
      Total number of images & 11,132,486 \\
      Total number of days&  75\\
      Storage  size&   2 Terabyte\\
      \hline
    \end{tabular}

\end{center}
\end{table}

\section{Conclusion}
\label{sec:conclusion}
In this paper we described the protocol and the tools that were developed to collect audio visual data of participants  with chronic conditions, in a simulated hospital environment.  
This data has served as a surrogate to in-patient data in the development of novel AI algorithms. The operational lessons that were learned during the first phase of the study will be integrated in a subsequent study to be conducted in the hospital. 
The main contributions of this paper are the protocol for an audio-visual collection study in a hospital and the software and hardware components that were created to support it as well as exposing design considerations and thought processes that have been considered while designing the study. 
The innovation of a cloud-native data collection study is also worth mentioning as it should allow the team to scale the study from a simulated hospital environment to a large scale hospital environment where hundreds of Terabytes of audio-visual and physiological data are expected while supporting a distributed team of AI researchers in a cost-effective manner. 
The data that has been collected in the study so far is already a unique and valuable resource allowing researchers to explore novel AI tools. 
As subsequent phases of the study become available and as the quality of the automated features extraction algorithms improves, the value of this data set is expected to grow. 

Going forward, two parallel tracks can be expected. First, the data that was already collected, would be progressively enriched with additional and improved features. It could also be enriched with manual tagging of SMEs.
Second, the study should be deployed in a real hospital. Once deployed, audio-visual and physiological data from patients would be used to validate and improve current algorithms. It is expected that these effort should allow the AI tools to accurately detect early stage patient deterioration, allowing them to become a useful tool in modern patient care.

\section*{Acknowledgment}
We would like to thank our study partners at \textit{Bold Insight}(https://boldinsight.com) who manged the study operation on behalf of GE Healthcare, including regulatory aspects with the Institutional Review Board (IRB), the patient recruitment specifics and data handling details. We to thank \textit{EarlySense}(https://www.earlysense.com/) for assisting us in setting up the physiological device  for the study. We also want to thank our manufacturing partners at \textit{MEC - Engineering Solutions} (http://www.mecad.co.il/) who were responsible for the design, development and manufacturing of the hardware and software for the data collection device.

\bibliographystyle{IEEEtran}
\bibliography{IEEEabrv,the_bibliography}

\end{document}